# OreProof: Verifiable Provenance with Limited Disclosure for Critical-Minerals Supply Chains Using Zero-Knowledge Proofs

Oleksandr Hrabar[1], Hossein Arshadi Soufiani[2], Henry M. Kim[2]*, Chien-Chih Chen[1], Ali Vazirizadeh[3], Hjalmar Turesson[2]
[1]University of Waterloo [2]Schulich School of Business, York University [3]Aisimpro

## Abstract

*Critical-minerals supply chains face a structural tension: regulators and buyers demand verifiable provenance, yet upstream actors are hesitant to disclose supplier identities, assay grades/yields, and prices that verification appears to require. We report a design science account of OreProof, a prototypical traceability platform addressing this verifiability-disclosure trade-off. Instantiated for gold, OreProof combines a hybrid on-chain/off-chain data model, Groth16 zero-knowledge proofs for selective disclosure, a Merkle-batched anchoring pipeline, and UNTP-aligned verifiable credentials on a public zkEVM testnet. Against a transparent baseline, directly inferable confidential attributes fell from three of four categories to none under a defined attacker model, while batched anchoring substantially improved throughput. Our contributions are the artifact prototype as well as four nascent design principles: prove over committed data rather than exposing it; credential only verifiable origin and flag unknown inputs for blended commodities; emit standards-aligned credentials from the outset; and partition disclosure by supply-chain role.*

**Keywords:** Blockchain Traceability; Zero-knowledge proofs; Selective Disclosure; Digital Product Passport; Critical Minerals

## 1. Introduction

The energy transition has made minerals such as gold, cobalt, lithium, nickel, and graphite strategically critical, and has intensified regulatory pressure to prove where they come from and under what conditions they were produced. The OECD Due Diligence Guidance for Responsible Supply Chains of Minerals from Conflict-Affected and High-Risk Areas (Organisation for Economic Co-operation and Development [OECD], 2016) establishes government-backed expectations for upstream-to-downstream due diligence, and sector frameworks such as the LBMA Responsible Gold Guidance (London Bullion Market Association [LBMA], 2021) make that guidance effectively mandatory for accredited refiners. In parallel, the European Union is operationalizing product-level transparency as a market-access condition through the Ecodesign for Sustainable Products Regulation (ESPR; European Parliament & Council, 2024) and the EU Battery Regulation 2023/1542 (European Parliament & Council, 2023), both of which mandate digital product passports carrying provenance and due-diligence data.

These regimes create structural tension though. A claim that gold is conflict-free, or that a battery's cobalt meets due-diligence criteria, is only trustworthy if independent parties can verify the evidence behind it. Yet disclosure is costly to upstream actors. Supplier lists, assay grades, transformation yields, and counterparty pricing constitute the data that confer competitive advantage, and that suppliers are most reluctant to surrender. Systematic reviews of blockchain traceability consistently identify this confidentiality concern as a principal barrier to adoption (Dasaklis et al., 2022; Rejeb et al., 2023). We call this the verifiability-disclosure trade-off. Disclose too little and verifiability collapses; disclose too much and supplier secrets leak. Most deployed blockchain traceability systems do not resolve this trade-off so much as pick a side. Fully transparent ledgers broadcast sensitive relationships (Malik et al., 2022), while permissioned ledgers with coarse access control hide so much that downstream verifiers must simply trust the upstream party, which defeats the purpose of traceability (Viguerie et al., 2024).

A second obstacle is standards misalignment. A platform that is technically sound but does not represent data in the format and vocabulary that emerging passport regimes expect imposes duplicate compliance cost and forecloses market access. The UN Transparency Protocol (UNTP), built on W3C Verifiable Credentials and Decentralized Identifiers (W3C, 2025) and advanced through UNECE Recommendation No. 49, Transparency at Scale (UNECE, 2024), is intended as the cross-border interoperability layer beneath national and sectoral passports. Aligning to UNTP and to the EU DPP early is therefore a design requirement, not an afterthought.

In this paper, we address a design question compelled by these obstacles: how can a blockchain traceability platform for critical minerals be designed so that it produces independently verifiable provenance claims while protecting upstream commercial confidentiality, and remains aligned with emerging digital product passport standards? To address this question, we prototype OreProof and discern general design principle from its operation.

*Corresponding author. hkim@schulich.yorku.ca

*An updated version of this paper appears in the Proceedings of the 60th Hawaii International Conference on Systems Science (HICSS-60), Honolulu, HI, January 5-8, 2027.*

We instantiate the design for a gold provenance use-case. The OreProof platform keeps confidential data off-chain under the owner's control and records only commitments, anchors, and verification references on the ledger. It uses Groth16 zero-knowledge proof (Groth, 2016) so an actor can prove a claim, such as approved origin or minimum purity, without revealing the underlying value. It also batches many off-chain lifecycle events into a small number of on-chain anchor transactions and issues UNTP-aligned verifiable credentials (UNECE, 2024; W3C, 2025) that EU passport regimes can consume directly. We deploy and benchmark the contracts on a public zero-knowledge Ethereum Virtual Machine (zkEVM) testnet.

Through the OreProof prototype, we aim to make three contributions: an instantiated artifact that addresses the verifiability-disclosure trade-off for critical-minerals traceability; a set of design principles for credentialing and selective disclosure in commodity supply chains, and a concrete standards-gap analysis for the minerals sector. We note that selective disclosure particularly benefits smaller participants, lowering the participation cost for small-scale and artisanal miners and SMEs that cannot expose their entire book to every counterparty. The remainder of the paper reviews related work and the standards landscape (Section 2), develops the verifiability-disclosure trade-off as kernel theory (Section 3), describes the system design and prototype (Section 4), reports the evaluation (Section 5), and discusses implications and limitations (Sections 6 and 7).

## 2. Literature Review

### 2.1 Privacy-preserving traceability systems

Blockchain has been widely proposed as an infrastructure for supply-chain traceability, but the resulting systems vary widely in how, and whether, they protect sensitive data. Systematic reviews catalogue blockchain traceability implementations across food, pharmaceutical, and agricultural settings and describe a maturing but largely proprietary research base (Dasaklis et al., 2022; Rejeb et al., 2023). A narrower stream tackles confidentiality directly. PrivChain attaches zero-knowledge range proofs and commitments (a published cryptographic value that binds to hidden data without revealing it) to provenance data so that actors prove claims without exposing trade secrets, but it operates on a single consortium ledger without standards alignment or rollup-based scalability (batching many off-chain operations into one on-chain commitment to raise throughput) (Malik et al., 2022). DECOUPLES provides unlinkability, preventing an observer from tying an actor to their transactions or linking separate events to the same party (El Maouchi et al., 2019). This hides actor identities but does not let a regulator verify a specific provenance claim, and it predates the passport regimes. Glew et al. (2022) adopt a design science method anchored on GS1 EPCIS but rely on data sanitization rather than cryptographic proof and are not minerals specific. These works advance one or two of the capabilities our problem requires, but no single system combines them.

Zero-knowledge proofs let a prover establish that a statement holds without revealing the data behind it, and succinct constructions such as Groth16 yield constant-size proofs with inexpensive verification well suited to on-chain checking (Groth, 2016). Applied to supply chains, this allows predicate claims (conditions that evaluate to true or false, like whether an attribute value lies within an approved threshold) to be checked against a commitment rather than disclosed. This is the cryptographic basis for the prove-do-not-expose approach the artifact operationalizes.

### 2.2 Digital product passports as an artifact

Recent IS work treats the digital product passport (DPP) as a design object rather than a purely regulatory artifact. Heeß et al. (2024) conceptualize DPPs for a low-carbon hydrogen market and find that stakeholders' willingness to share data hinges on addressing privacy and disclosure concerns, the same tension our design targets. Illán García et al. (2024) implement DPP management with decentralized identifiers and verifiable credentials supporting selective disclosure under the ESPR, but their system is product-generic and does not report UNTP alignment or a zk-rollup execution environment. The GS1 EPCIS 2.0 event model contributes a complementary vocabulary, including the TransformationEvent type that represents inputs being consumed into new outputs, as required to model bulk-mineral refining (GS1, 2022).

### 2.3 Enterprise blockchain governance and architecture

Enterprise blockchains are typically permissioned and, despite decentralization rhetoric, often reproduce centralized control; Viguerie et al. (2024) theorize Chief, Clan, Custodian, and Consortium governance archetypes shaped by participant dominance and platform openness. At the architectural level, the dominant pattern stores sensitive payloads off-chain while committing only hashes or references on-chain, reconciling integrity with confidentiality and data-protection constraints. Our artifact adopts this pattern and extends it with commitment-based selective disclosure.

To our knowledge, no peer-reviewed system unites a hybrid on-chain/off-chain architecture, a zk-rollup execution environment, zero-knowledge selective disclosure, and UNTP and EU DPP alignment for critical-minerals traceability. The comparables reviewed so far each contribute provenance with privacy, decentralized unlinkability, or standards-aligned credentials, but not the combination. The efforts that touch all of these at once are non-peer-reviewed industry pilots, such as Circulor's blockchain tracing of cobalt from the Democratic Republic of the Congo (Soerensen, 2024) and the Google-led tin traceability pilot with Minsur and Minespider in Peru and Rwanda (Google, 2020). Our work addresses that gap.

## 3. Theoretical Framing: The Verifiability-Disclosure Trade-off

Operations management literature treats transparency, visibility, and traceability as central supply-chain concerns and has examined both the design of blockchain traceability and the disclosure dilemma it creates, namely that the information sharing which improves coordination also exposes participants to competitive harm (Cui et al., 2024; Hastig & Sodhi, 2020). This tension is acute in extractive settings such as cobalt mining (Hastig & Sodhi, 2020). A recurring theme runs through extant literature: end-to-end traceability requires upstream actors to contribute operational data that is commercially sensitive, so the evidence most needed for verification is the data suppliers are least willing to expose. We take this tension as the "***kernel theory***[1]" guiding the design and state it as a trade-off between two constructs: Verifiability and Confidentiality.

Verifiability (V) is the degree to which an independent party can confirm a provenance or conformity claim from available evidence. Confidentiality (C) is the degree to which commercially sensitive attributes, such as counterparty identities, grades, yields, and prices, are protected from unauthorized parties. Conventional traceability designs treat V and C as substitutes along a single dial: disclosing more evidence raises V and lowers C, while disclosing less protects C at the cost of V. Under this view the two cannot improve together, which is why transparent ledgers leak supplier data and access-controlled ledgers collapse independent verification.

The insight that motivates our OreProof design is that verifiability depends on whether a claim can be checked, not on whether the data grounding the claim is disclosed. A zero-knowledge proof lets a prover establish that a committed value satisfies a stated predicate without revealing the value (Groth, 2016); this decouples V from C, so a proof can sustain V while C is held fixed, shifting the achievable region outward rather than sliding along the substitution frontier (Figure 1). Prior work has shown that proofs over committed provenance data can preserve confidentiality while supporting verification on a ledger (Malik et al., 2022); our framing generalizes this into a design objective. The resulting meta-requirement is that each actor discloses only the minimum needed for a counterparty or regulator to verify a specific claim, and nothing more.

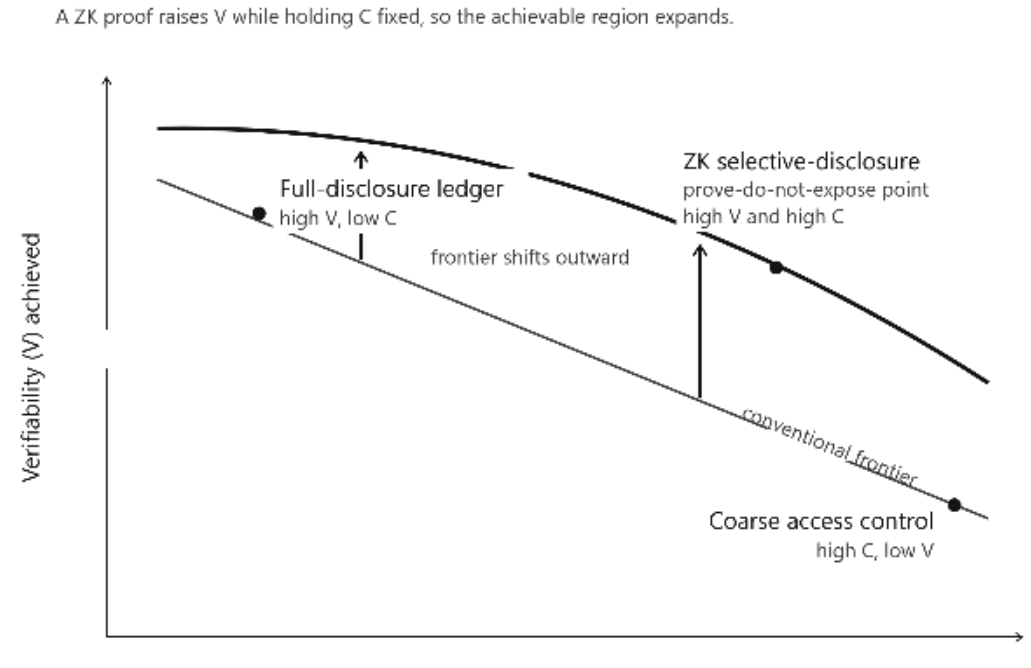

**Figure 1.** The verifiability-disclosure frontier. Conventional designs trade V against C along a single substitution frontier. Zero-knowledge selective disclosure raises verifiability while holding confidentiality fixed, expanding the achievable region toward a prove-do-not-expose operating point.

The "specific claim" from Section 3 is instantiated as two bounded predicates relevant to gold attestation: approved-origin set membership, and minimum grade and purity thresholds. Each is proved against an on-chain commitment, so an actor can establish that a claim holds without revealing the underlying value.

Confidentiality is operationalized as privacy leakage, the fraction of confidential attributes an outside party could infer from public on-chain data. The framing predicts that confidentiality can be held fixed while claims remain verifiable, a prediction the evaluation tests directly (Section 5). We are careful not to overstate the construct: verifiability is demonstrated for the implemented predicates rather than measured as a general quantity, and any privacy result is leakage under a specific attacker model rather than a guarantee of absolute privacy.

[1] **Kernel theory.** Conventional designs treat verifiability (V) and confidentiality (C) as substitutes; zero-knowledge selective disclosure decouples them, allowing V to be sustained while C is held fixed.

# 4. System Design and Architecture

## 4.1 Overview and design rationale

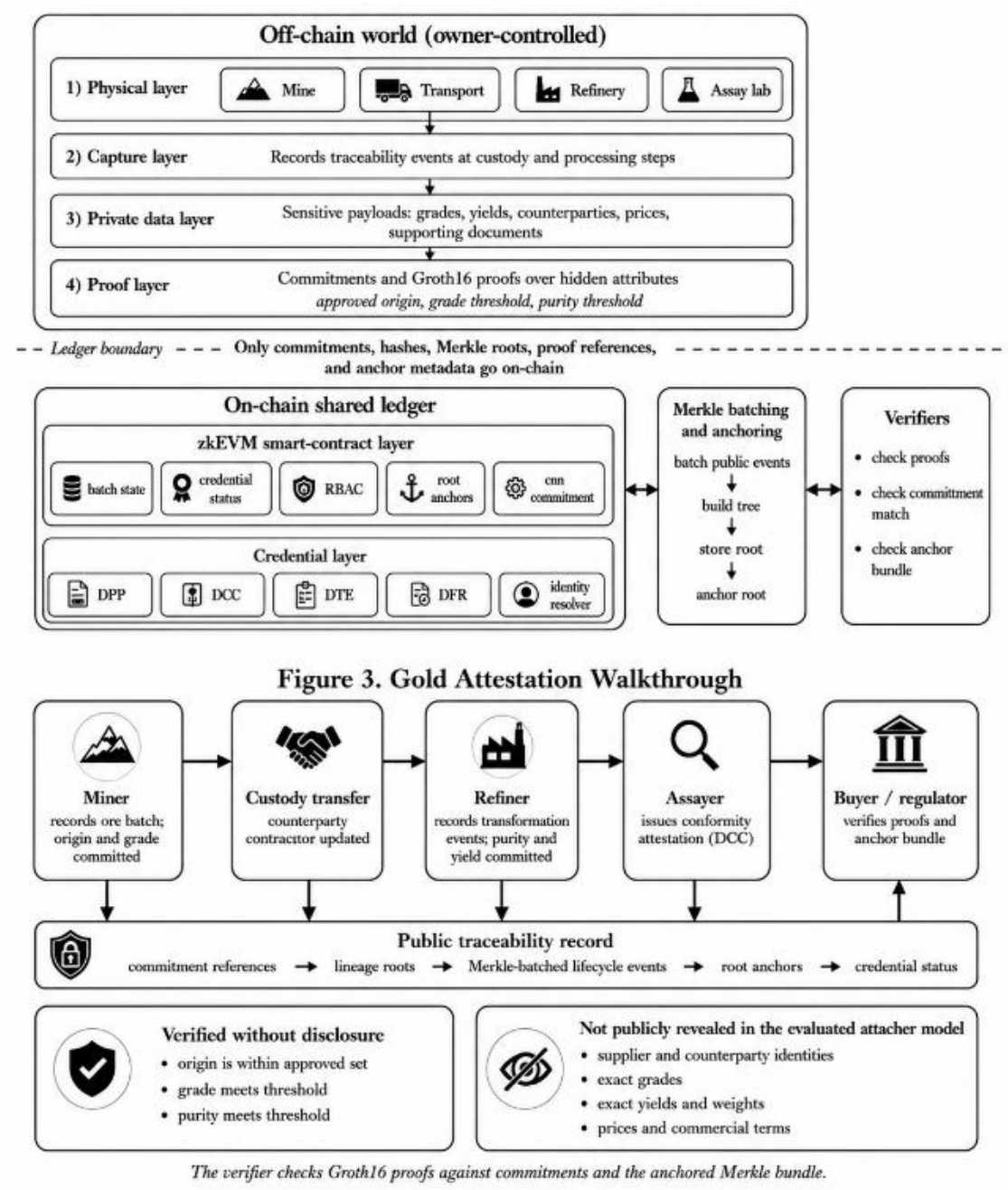

**Figure 2.** Reference architecture of the platform. Traceability data remains off-chain, while ZK proofs, commitments, and UNTP-aligned credentials are managed through zkEVM smart contracts. Only commitments, hashes, and proofs are recorded on the shared ledger.

The platform is a permissioned, hybrid on-chain/off-chain traceability system for critical-minerals supply chains, instantiated in a gold provenance setting, designed to reconcile verifiable provenance with the protection of commercially sensitive upstream data. Confidential attributes, including counterparty identities, exact assay values, transformation yields, prices, and commercial terms, are retained off-chain under application control, while the blockchain records only audit-oriented artifacts such as commitments, Merkle-root anchors, identifier metadata, and verification references. This separation preserves verifiability while avoiding full disclosure of operational business data, consistent with the common enterprise pattern of keeping sensitive payloads off the shared ledger (Section 2).

Five design decisions follow from the "***kernel theory***" (that verifiability need not be traded against confidentiality) and the problem context (Figure 2): a hybrid data model, a zk-rollup execution environment, zero-knowledge selective disclosure, role-based access for miner, refiner, assayer roles consistent with the EU and UNTP mandates, and standards-aligned credential output. The contracts were deployed and validated on a public zk-rollup testnet (Polygon zkEVM Amoy); we treat the specific chain as a substitutable component.

## 4.2 Smart-contract layer: two lifecycle models

The smart-contract layer uses an initial direct-write lifecycle model as a transparent baseline: it records lifecycle events explicitly on-chain but exposes more information than the paper's privacy claims permit. Our design affords an additional privacy-preserving approach. Rather than writing event data to the chain, an actor posts a cryptographic commitment to the confidential data. The actor also anchors a batch of off-chain events using a single Merkle root, one hash that stands in for the whole batch and lets any individual event be verified later without exposing the rest. The confidential data never leaves the actor's off-chain systems. Under this approach the blockchain acts as a verifiability layer rather than a public record of business state.

## 4.3 Off-chain lifecycle logic and batched anchoring

The main traceability logic is implemented off-chain in an application service that manages the lifecycle of a mineral batch across four event types, each originating at one of the actors in Figure 2 physical layer: ore registration (at the mine), custody transfer (during transport), transformation or refining (at the refinery), and certification or attestation (at the assay lab). Each event is converted into a canonical record, inserted into a Merkle batching pipeline, and later anchored on-chain by submitting only the root and the associated scope metadata (e.g. batchID, eventType, actorRole, actorID, merkleRoot), so that many lifecycle events are finalized through a small number of root-anchor transactions. This realizes the design principle that the chain stores proof-oriented data rather than confidential operational payloads. For bulk minerals, which are physically blended, the service applies a conservative blending model: where input provenance is uncertain, that uncertainty is recorded explicitly rather than converted into an unsupported origin claim. This is a deliberate design choice against overclaiming that mass-balance accounting can invite.

Selective disclosure is implemented with Groth16 zero-knowledge proofs (Groth, 2016). For the gold attestation scenario (Figure 3), the prototype supports predicates for approved-origin set membership and minimum grade and purity thresholds. In each case the data owner commits to a confidential value when the corresponding batch attribute is recorded, and later proves that the value satisfies the required predicate without revealing the value itself; a verifier, such as a downstream buyer, a conformity assessment body, or a regulator, checks the proof against the corresponding on-chain commitment. This allows a specific claim to be verified without exposing the underlying commercial data, operationalizing the prove-do-not-expose meta-requirement ("each actor discloses only the minimum needed for a counterparty or regulator to verify a specific claim, and nothing more.").

**Table 1.** Mapping of platform objects to UNTP/EU DPP and identified standards gaps.

| Platform object | UNTP credential / EPCIS event | EU DPP touchpoint | Identified gap |
|---|---|---|---|
| Batch provenance record | Digital Product Passport | DPP material provenance | Batch-vs-shipment granularity: DPP granularity assumes a serialized item or shipment, not the batch granularity bulk minerals require |
| Refining step | Digital Traceability Event / TransformationEvent | Lifecycle data | Mining-specific event types (assay, smelter-charge, refining-transformation) are not natively modeled |
| Assay / conformity attestation | Digital Conformity Credential | Due-diligence evidence | Multi-entity Canadian conformity chains: DCC assumes a single conformity assessment body, not chained multi-entity attestations |
| Actor / facility identity | Digital Facility Record / Identity Resolver | Economic-operator identity | Canadian regulatory identifiers are absent from the identifier scheme register |
| Selective redaction | Decentralized Access Control | Role-based access | Field-level selective redaction is needed beyond document-level access |

## 4.4 Standards alignment and credential mapping

Standards alignment is provided through UNTP-oriented credential and discovery components. The platform supports decentralized-identifier resolution, conformity-credential linkage, and generation of standards-aligned verifiable credentials (UNECE, 2024; W3C, 2025), and it maps its internal objects to UNTP credential types and EU DPP data requirements, with the GS1 EPCIS TransformationEvent used to represent the consumption of inputs into outputs (GS1, 2022). The prototype therefore operates not only as a provenance tracker but also as a credentialing system: blockchain anchors provide tamper-evident references, while higher-level credentials express origin, conformity, and product-passport semantics. Table 1 summarizes this mapping together with the standards gaps identified during a public-review gap analysis, which we discuss in Section 6.

The architecture is assessed in Section 5 against the study's two key performance indicators, throughput and privacy leakage. Throughput is defined as sustained committed business-event throughput under the batching architecture rather than an isolated smart-contract call rate, and privacy leakage as the fraction of confidential attribute instances inferable from public on-chain and benchmark-visible data. These definitions ensure the evaluation corresponds to the hybrid, zk-enabled, batched-root architecture described here rather than to a fully transparent on-chain registry. All measurements are reported in Section 5.

# 5. Evaluation

We evaluate our privacy-preserving architecture against the study's two key performance indicators, throughput and privacy leakage, using the initial direct-write model as a transparent baseline. Both prototypes are evaluated on the same live testnet deployment (Polygon zkEVM Amoy), so the comparison isolates the effect of moving from a fully transparent on-chain state machine architecture to a hybrid, zk-enabled, batched-root architecture rather than any difference in underlying infrastructure.

## 5.1 Throughput

The two architectures are measured at different levels, transaction-level for the baseline and batch-level for the privacy-preserving design, and distinction matters for interpretation. In the baseline, lifecycle events are recorded by a fully transparent on-chain state machine in which ore registration, refining, and certification are each a separate transaction, so one complete business flow requires three sequential on-chain writes (registerOre, refine, certify). Throughput was the end-to-end committed transaction rate, $T_{legacy} = N_{tx} / \Delta t$[2], where $N_{tx}$ is the number of mined lifecycle transactions and $\Delta t$ is the wall-clock time from submission to final confirmation. Under sequential lifecycle dependencies, the live Amoy deployment committed 0.665 transactions per second, which reflects the prototype's practical execution rate rather than the chain's theoretical capacity.

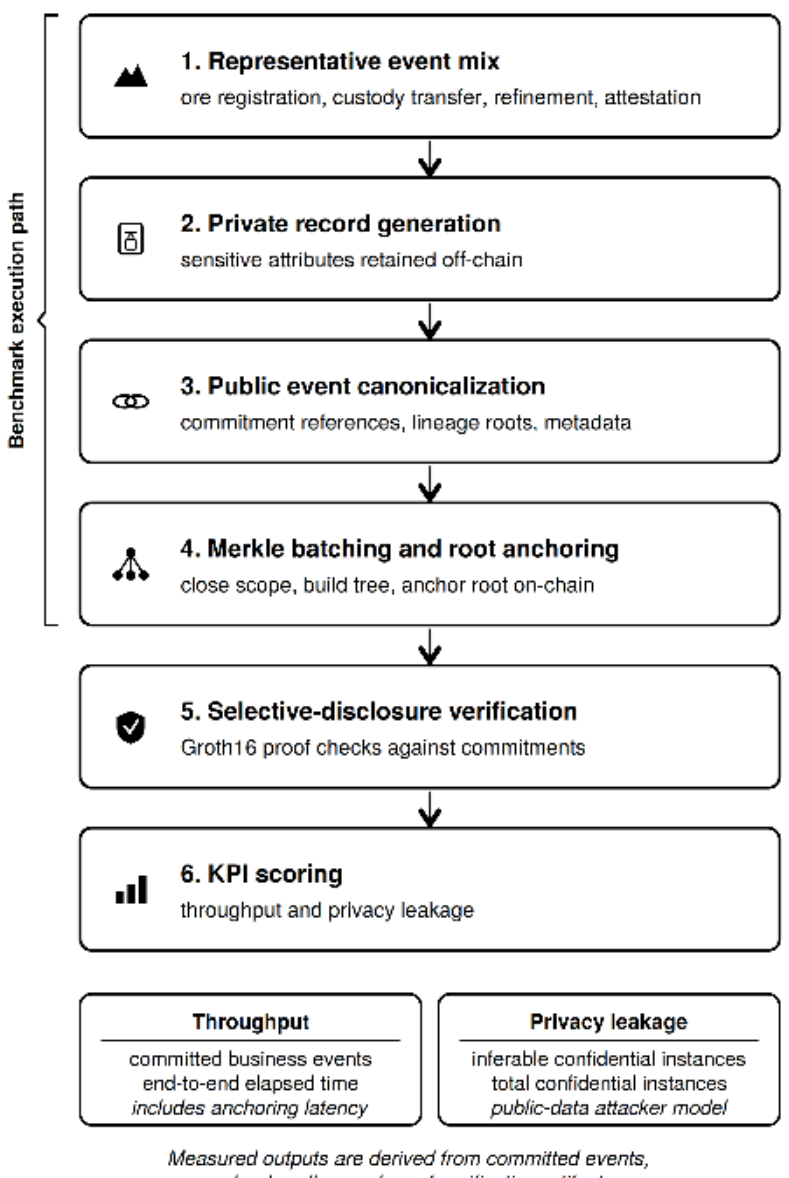

**Figure 3.** Evaluation and Measurement Workflow

The privacy-preserving architecture measures throughput at the business-event level, $T = N_{events} / \Delta t$, where $N_{events}$ is the number of private lifecycle events accepted into the off-chain pipeline and durably committed through Merkle-root anchoring, and $\Delta t$ is the elapsed time from ingestion start to final anchor confirmation. This reflects that the system no longer writes every event as its own transaction but instead finalizes many events through a smaller number of root-anchor transactions. In the live Amoy benchmark, the prototype committed 2,000 private events using four Merkle-root anchor transactions at a batch size of 500, with a total end-to-end time of 14.368 s, comprising 0.725 s of off-chain ingestion and 13.643 s of on-chain anchor confirmation. The resulting committed business-event throughput was 2000 / 14.368 = 139.2 events/s. This is durable committed throughput rather than the event-creation rate: off-chain ingestion alone reached 2,759 events/s, but the reported figure is lower because events are counted only after all anchor transactions confirm, making the anchoring phase, not local ingestion, the dominant cost.

To avoid conflating different rates, we distinguish them explicitly. The baseline commits 0.665 transactions per second with on-chain durability. Since each transaction carries one lifecycle event, this is also about 0.665 committed events per second. The redesigned system ingests off-chain events at up to 2,759 events per second, but this figure excludes on-chain durability and is therefore not comparable to the baseline. The comparable quantity is committed throughput — events that are durably anchored — which is 139 events per second. On a reasonable like-for-like basis (committed events per second), the privacy-preserving architecture has substantially higher throughput[3] than the baseline, because batching amortizes one anchor transaction over many events. The baseline's advantage lies elsewhere: each event receives independent on-chain finality immediately, whereas the redesigned system defers finality until its batch anchors, trading per-event latency for aggregate committed throughput. This trade-off, not raw speed, is the substantive difference between the two designs.

## 5.2 Privacy leakage

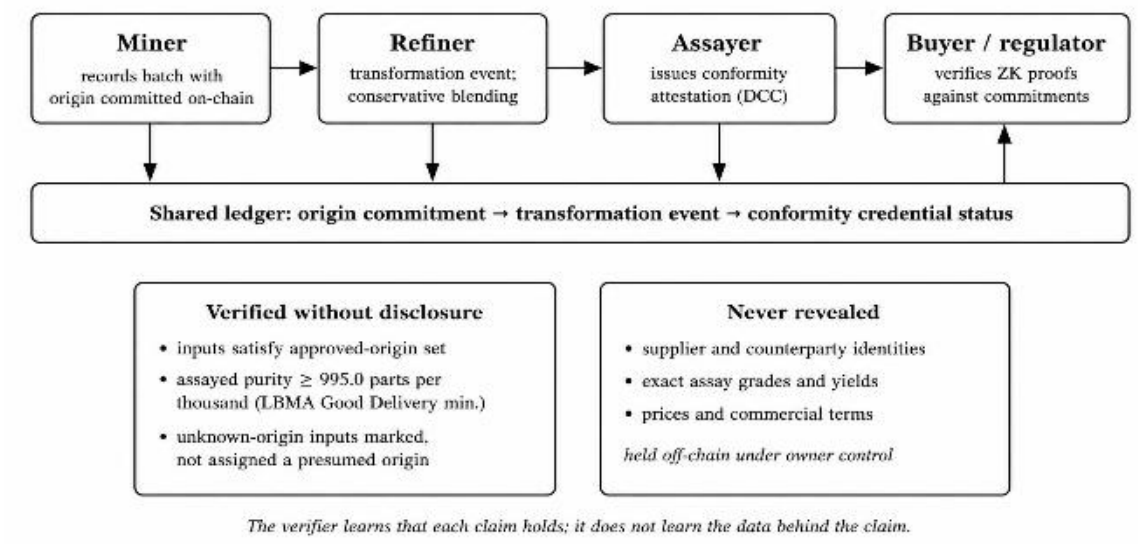

**Figure 4.** Each actor logs commitments and events on the shared ledger, and the verifier uses ZK proofs to confirm origin and purity without revealing supplier identities, grades, yields, or prices.

Privacy leakage is the proportion of confidential attribute categories inferable from public, benchmark-visible data, $L = (N_{inferable} / N_{confidential}) \times 100$, evaluated under a defined attacker model. The attacker is assumed to see only public article-mode benchmark objects, namely on-chain commitments, Merkle roots, anchor metadata, and benchmark-visible outputs, and a category is counted as leaked only when a confidential field is directly readable or deterministically derivable from those objects.

---

[2] The subscript 'legacy' denotes the initial direct-write model, which serves as the transparent baseline throughout; we use 'baseline' and 'legacy' interchangeably. The redesigned privacy-preserving architecture is denoted 'ZK-hybrid'.

[3] Though 139.2 >> 0.665, the two architectures are sufficiently different that rather than stating the difference as a numerical factor (i.e. 209 times higher throughput), a qualitative descriptor like “substantially higher throughput” is the appropriate characterization.

Four confidential categories are considered: counterparty identity, grade or purity, transformation yield, and price. In the baseline, counterparty identity is inferable from public custodian addresses, grade or purity from public assay fields, and transformation yield from public input and output weights, while price is not exposed. Three of four categories are therefore inferable, giving $L_{legacy} = (3 / 4) \times 100 = 75\%$. In the privacy-preserving architecture, public objects are limited to commitments, hashes, Merkle roots, and anchor metadata; counterparty identities, exact grade values, yields, and prices are not published in plaintext, and none of the four categories are inferable, giving $L_{zk\text{-}hybrid} = (0 / N_{confidential}) \times 100 = 0\%$, which satisfies OreProof development team's internal privacy-leakage target of below 0.1 percent under this model.

The **0%** figure must be read narrowly. It is observed direct attribute leakage under the benchmark attacker model, not a claim of absolute system-wide privacy, and it reflects only confidential fields that are directly readable or deterministically derivable from the public on-chain data checked. It therefore excludes residual risks that fall outside this model, including metadata correlation, timing or linkage across repeated events, inference from batch sizes or shipment patterns, weak-salt guessing against small-domain commitments, auxiliary external knowledge, and off-chain database compromise. The reduction from **75%** to **0%** is directly attributable to the use of commitments and zero-knowledge selective disclosure.

The two gains come from different mechanisms: zero-knowledge selective disclosure removes the directly inferable attribute leakage, while batched anchoring delivers the higher committed throughput. The contribution is not that zero-knowledge proofs make transactions faster, but that they enable a private-verifiability model which, combined with batched anchoring, yields a higher-throughput, more scalable traceability architecture.

## 6. Discussion

OreProof's development yields transferable design knowledge alongside implications for practice and standardization. We abstract four nascent design principles, each derived from a specific design decision and supported by the evaluation or the kernel theory of Section 3. They are nascent in the design-science sense: grounded in a single instantiation and offered as transferable guidance to be tested through use, not as validated laws. DP1 is supported directly by the privacy-leakage result; DP2 through DP4 are design stances whose general value requires further instantiation and naturalistic (Venable, Pries-Heje and Baskerville, 2006) evaluation.

**DP1 (Prove, do not expose):** to reconcile verifiability with confidentiality, design the system so that actors submit proofs over committed data rather than the data itself. This is the design expression of the kernel theory, and the principle the privacy-leakage result most directly supports.

**DP2 (Conservative claiming for blended commodities):** for bulk materials that are physically blended, credential only what can be proven and mark unknown inputs explicitly, rather than propagating probabilistic origin claims. The system claims less but proves what it claims: it credits only verifiable origin and flags unknown inputs, rather than averaging blended material into a single origin claim that the evidence does not support.

**DP3 (Standards-adhering output):** emit traceability data as standards-aligned verifiable credentials from the outset, so that compliance and interoperability are properties of the artifact rather than later bolt-ons.

**DP4 (Role-structured disclosure):** partition disclosure by supply-chain role so each party receives the minimum view required for its function, consistent with the role-based access models in the EU Battery Regulation and UNTP (European Parliament & Council, 2023; UNECE, 2024).

Implementing DP3 requires mapping the platform's objects onto UNTP and EU DPP credential types, and that mapping surfaced four gaps that matter beyond this project (Table 1). First, the standard event taxonomies have no built-in types for mining-specific steps such as assaying, smelter-charging, and refining. Second, passport models assume a serialized item or a discrete shipment, whereas bulk minerals move as batches that are blended and split, so the granularity does not match. Third, the standard conformity credential assumes a single assessment body, but mineral chains involve several parties each attesting to part of a claim. Fourth, national regulatory identifiers are not yet covered by the standard identifier schemes. Because these standards are still being finalized (UNECE, 2024; GS1, 2022), we offer these four gaps as concrete input to UNTP and EU DPP standardization for the minerals sector, not as permanent shortcomings of the standards.

For enterprise integration, the hybrid on-chain/off-chain design, combined with zero-knowledge selective disclosure, lets firms participate in shared traceability without surrendering trade secrets, addressing the confidentiality concern that the literature identifies as a principal barrier to traceability adoption (Hastig & Sodhi, 2020; Dasaklis et al., 2022). Standards-aligned output turns compliance into market access, since the data firms must produce for the EU DPP and battery-passport regimes doubles as the credential that lets them trade; and selective disclosure lowers the cost of participating by letting each actor reveal only what a claim requires, which helps SMEs and small-scale miners who do not wish to expose their whole position to every buyer. These are design-level arguments grounded in OreProof development and the cited literature; quantifying the adoption or cost effects would require field deployment, which we leave to future work.

Finally, selective disclosure can lower participation barriers for small-scale and artisanal producers, and it offers a mechanism to honor confidentiality where provenance data intersects Indigenous land rights or differs in sensitivity across jurisdictions. These benefits depend on governance choices about who may issue and verify credentials, and enterprise blockchains frequently reproduce centralized control rather than the decentralization their architecture suggests (Viguerie et al., 2024). Whether selective disclosure advances equity and fairness therefore turns on the governance model layered above it, which we treat as future work rather than a property that our artifact guarantees.

## 7. Concluding Remarks

We present OreProof, a proof-of-concept hybrid, zk-enabled, UNTP-aligned traceability platform for critical minerals, instantiated for gold, that addresses the verifiability-disclosure trade-off by letting actors prove provenance and conformity claims without exposing the sensitive data behind them. Against a transparent direct-write baseline, the redesigned architecture reduces directly inferable confidential attributes from three of four categories to none under the benchmark attacker model and substantially increases committed throughput under batched anchoring. Privacy gain is attributable to commitments and zero-knowledge selective disclosure, and the throughput gain to the batching architecture rather than the proofs themselves. Our research contribution is the instantiated artifact, four nascent design principles for credentialing and selective disclosure in commodity supply chains, and a standards-gap analysis for the minerals sector. Because the blockchain serves only as a substitutable verifiability layer, and the design (commitments, selective-disclosure predicates, batched anchoring, and standards-aligned credentials) is independent of both the specific chain and the gold scenario, the prototype is best read as a proof of concept that can be migrated to another execution environment and mapped to other minerals and conformity regimes by substituting predicates and credential mappings rather than rebuilding the architecture.

Several limitations bound these claims. The artifact is implemented on a public zkEVM testnet rather than in production; the privacy figure is observed direct leakage under a specific attacker model rather than a guarantee of absolute privacy; the throughput figure counts committed business events rather than raw transactions; and whether a formal LBMA-recognized attestation can be issued through this mechanism remains open. We are also aware that the blockchain used, the Polygon zkEVM Mainnet Beta sequencer, is scheduled to be sunset on 1 July 2026. This makes a migration or chain-selection decision necessary and is why we treat our specific chain source-code on that testnet as replaceable rather than integral to our contribution. Moreover, our design principles are nascent and require further instantiation and more rigorous evaluation to mature toward design theory. Future work includes production deployment, evaluation in industrial partners' production environments, extension to additional minerals, and engagement with standards bodies on the identified gaps. By treating standards-aligned, privacy-preserving verifiability as a design objective rather than an unavoidable trade-off, our OreProof platform offers a path for critical-minerals supply chains to meet rising regulatory and market demands for provenance without forcing upstream actors to choose between compliance and confidentiality.